\documentstyle[prb,aps,twocolumn,amssymb,amsfonts,epsfig]{revtex}

\begin{document}
\draft \flushbottom

\wideabs{

\title{
Spinwaves and large electron-phonon coupling near the metal-insulator
transition in hole-doped high-T$_C$ oxides.
}

\author{T. Jarlborg}
\address{
DPMC, University of Geneva, 
24 Quai Ernest-Ansermet, CH-1211 Gen\`eve 4, Switzerland
}

\date{\today}

\maketitle

\begin{abstract}
A theory for spin and lattice couplings to the electronic states in high-T$_C$ oxides is presented, 
with HgBa$_2$CuO$_4$ as
an example. A simple analytical model
suggests that the barrel band is sensitive to potential perturbations with long-range Fourier components, and that
gaps can be opened at the Fermi energy. This is 
confirmed in self-consistent band calculations for modulations in elongated supercells, where spin waves
and phonon coupling compete for equal q-vectors.
It is argued that the wavelengths of spin-wave 
modulations and of the phonon modes with large coupling, depend on the doping.  
This mechanism supports the idea that  pseudogaps are caused by stripe like spin-modulations in underdoped systems,
while superconductivity is attributed to enhanced
electron-phonon coupling for long wavelength phonons. 

\end{abstract}


}

\begin{figure}[tb!]
\leavevmode\begin{center}\epsfxsize6.6cm\epsfbox{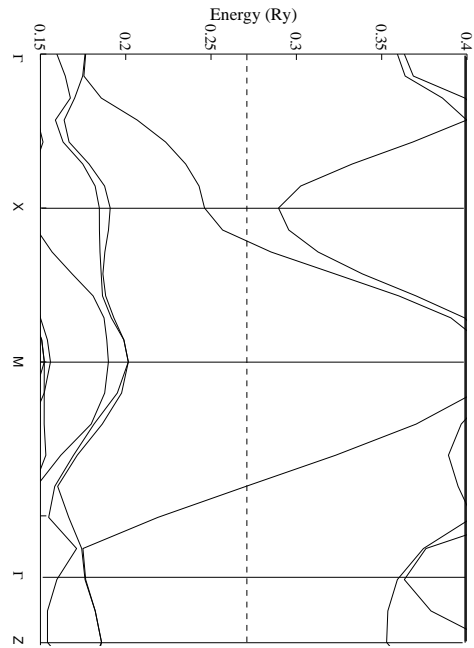}\end{center}
\caption{\label{figband}
Band structure of HgBa$_2$CuO$_4$ along symmetry directions in the $k_z$=0 plane of the tetragonal BZ.
The Fermi energy is indicated by the broken line. 
}
\end{figure}

The mechanism behind high-T$_C$ superconductivity in copper oxides is still unknown.  
From the vast amount of experimental information that has been collected from
high-T$_C$ oxides, one can recall some facts that seem to distinguish these materials
from ordinary low-T$_C$ superconductors \cite{oren,tim,tallon}. 
Phase sensitive tests of the
pairing indicate d-wave symmetry of the superconducting gap  \cite{tsuei}. 
This is also implied from
photoemission and tunneling measurements where the gap is highly anisotropic
over the Fermi Surface (FS) \cite{norm,renn}. Such measurements detect pseudogaps,
which extend well above T$_C$  in underdoped systems. They
appear to have the same symmetry as the superconducting gap, making the FS visible only along the
diagonal. The isotope effect, which is small 
and variable, exists also for the pseudogap \cite{hof}. Spin fluctuations follow the same type
of antiferromagnetic (AFM) orientation of ordered Cu moments that can be seen in undoped, insulating copper
oxides. Separation of metallic and insulating regions form static stripes for some compositions
\cite{tran}, but they might be dynamic in other cases. Band theory based on the density functional
approximation fails to describe the undoped, insulating cuprates, while the bands and FS of metallic, 
superconducting cuprates agree well with experimental data.
The electron-phonon coupling ($\lambda$) calculated from band theory \cite{kra,san}, is quite large in view of the 
low density-of-states (DOS) in these materials, but still not high enough
to explain the high T$_C$'s.
Here we perform self-consistent
electronic structure calculations based on the local (spin) density
approximation (LDA) \cite{lda}, for long wavelength excitations of either spin fluctuations or phonons, by use
of large supercells. As will be shown, spin waves and phonons seem to compete for the same q-vectors,
and the perturbations will be large at the FS, leading to a tendency for gap formation in both cases.
We focus on HgBa$_2$CuO$_4$ (HBCO), since it is relatively simple with one CuO-plane and only
one cylindrical FS, a so-called 'barrel'. Such a piece of FS is common to all high-$T_C$ oxides and probably
involved in the mechanism for superconductivity. The lattice constant $a$ is 7.32 a.u.. The bands are calculated using the 
linear muffin-tin orbital method, where
5 empty spheres are introduced to account for non-spherical potential corrections in the most open
parts of the structure. The structure, details of the method and  band results for the
basic unit cell, are described earlier \cite{bar,san}. Only Cu, and to some extent the planar
O, have a significant DOS at $E_F$. 
The band structure of the normal cell is shown
in fig. \ref{figband} for k-points in the $k_z$=0 plane. 
The FS is a cylindrical 'barrel' centered at the M-point.  The FS
reaches the zone boundary at X for a doping of about 0.25 hole per formula unit (f.u.).
The cell of the AFM insulating case, has two Cu in the xy-plane with opposite
magnetic moments. This cell of
two f.u. has 26 sites (16 atoms and 10 empty spheres).
The AFM state can be reached by introducing large correlation, increased exchange or by
application of external field. Here we include a staggered field (positive and negative with maximum amplitude of
30 mRy) on the two Cu in the 
AFM cell. The undoped material is insulating with this field, with a gap of 10 mRy and local moments of 
about 0.5 $\mu_B$. 
The critical field for having a gap is about 20 mRy.
 A uniform field within each Cu-sphere 
makes the entire Cu-bands to be shifted, so that part of the upper DOS-edge of the Cu-d band 
will get closer to $E_F$ with increasing field.

The 26-site cells are put together in the  [1,1,0]-direction, to form larger cells
in order to study q-waves modulated along the diagonal (with respect to the Cu-O bonding).
 Four supercells containing 104, 130, 156 and 208
sites are considered, corresponding to a maximal wavelength of $8 \sqrt{2} a$. 
For modulations along  [1,0,0] it necessary to have a two-layer wide unit cell (along y) because of
the underlying AFM order. A cell of 208 sites will describe spin modulation
with wavelength $8a$ along x. For studies of phonon modulations without consideration
of spins, we can use narrow cells (width $a$ along y), so that a 104-site cell can describe a modulation
with period [$8a$,0,0]. In each case we calculate the bands in 24 k-points in the respective irreducible 
Brillouin Zones (BZ). The DOS is determined by from tetrahedron k-point integration, where
it is assumed that the energies vary linearly within each tetrahedron. The use of few k-points may
introduce artifacts in the DOS, especially near gaps where the bands deviate from linearity.

The z-dispersion of the barrel band is very small,
but the dispersion is large in the xy-plane, as along the diagonal $\Gamma - M$ (cf. fig. \ref{figband}).
 A modulation of the potential $V_q \cdot$ sin($qr$), 
expressed by the Fourier coefficient
$V_q$, where q is along [1,1,0], will open a gap for a free electron like band. From the well-known solution
for bands near the zone boundary in semiconductors \cite{ziman},

\begin{equation}
E(k)= \frac{1}{2}(\epsilon_k+\epsilon_{k-q}) \pm \frac{1}{2}\sqrt{(\epsilon_k-\epsilon_{k-q})^2+4 V_q^2}
\label{eqek}
\end{equation}

which at the zone boundary $k=\frac{1}{2}q$, becomes $E(q)=\epsilon_{\frac{1}{2}q} \pm V_q$. Thus, there are
two states with a gap, $2 V_q$, in between, at the new BZ-limit at $\frac{1}{2}q$. The square of the wave function 
belonging to the state below the gap, has a modulation  in phase with the potential
modulation so that its amplitude is largest where the potential is most attractive.
 The  square of the wave function for the state above the gap has the same
modulation, but its phase is shifted, so that the large amplitude coincides with the most repulsive part of the 
potential. 

Consider first AFM cases. The
 AFM state in the undoped case is partly stabilized because of the opening of the gap at $E_F$.
Hole doping made in a rigid-band manner on the same DOS lead to metallic states, without the stabilizing
effect of the gap.  As will be seen, modulations of the AFM moments will open 
gaps at different doping levels, with possible energy gains. 
 An example of a modulation is when the Cu sites projected on
the [1,1,0] direction have moments; zero, up, down, up, zero, down, up, down. Such a "stripe" has 
 $\pm V_q$ components in the spin-potentials, 
with a wavelength covering totally 8 Cu layers.
If  $E_F$ is far below (or far above) the gap, both states will be occupied (or unoccupied) 
with no gain in total energy.  It is only if $E_F$
is near the gap that there will
be a possibility to gain total energy. The occupied state reduce its kinetic energy, while the raise in energy of
the upper, empty state does not contribute to the total energy.  The process is self-supporting via 
the exchange potential, since the increased spin density in regions where the potential is spin splitted
 will increase the potential splitting further. 
Thus, the argument is that the CuO-plane 
systems can chose the wavelength of the
modulation according to its doping to make an optimal gain in total energy. 
This mechanism always opens the gap (or pseudogap)
close to $E_F$, leading to a correlation between  
the wavelength and doping. These arguments can be used
as long as there are no other bands near $E_F$, because they would normally involve an increase of the
total energy when a modulation is imposed.  

Similar arguments can be used to discuss the coupling between phonon modes and the electronic states near $E_F$. 
For example, a phonon induced modulation of the
 potential on Cu sites will also have its 
Fourier component $V_q$. (Since the local Cu-DOS is the highest near $E_F$, it will be largest effects
from potential changes on Cu-sites).
The (non spin-polarized) bands will open a gap as for each of the spin bands
in the AFM case, if the gap is close to $E_F$. As before,
only one particular wavelength will be concerned, depending on the doping. It is worth noting
that for a given doping, the same q-vector is involved for both spin waves and phonons. 
This mechanism will be dynamic for phonons, unless the gain in energy will make the phonon completely soft to
make a lattice instability. Spin waves could be pinned to impurities and therefore be static, but coupling
between spin waves and phonons are likely since they compete for the same q-vectors.

\begin{figure}[tb!] 
\leavevmode\begin{center}\epsfxsize8.6cm\epsfbox{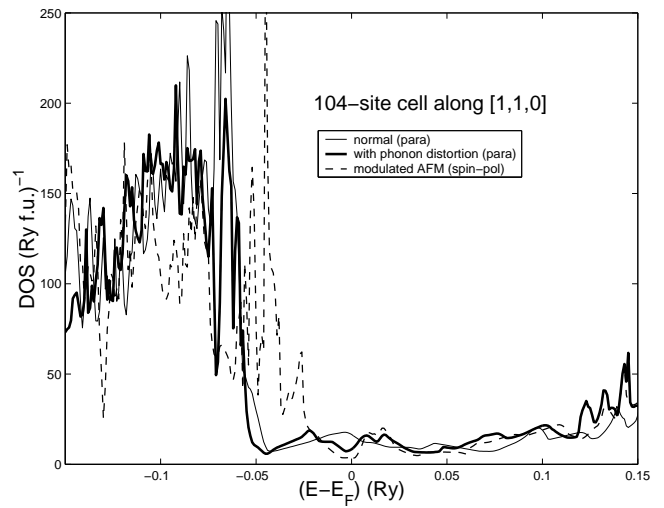}\end{center}
\caption{\label{dos104}
Paramagnetic DOS for a 104-site unitcell of HBCO (doped with 0.25 holes per f.u.)
 along [1,1,0] (thin line), paramagnetic DOS when the structure contains
 a sinus modulation of
Ba$_z$ positions (heavy line) and the DOS for a sinus-modulated AFM state along the cell (broken line). 
The latter is obtained via applied fields
(maximum amplitude 30 mRy) on the Cu sites. A higher field will open the gap completely, but also move the
Cu-d band edge closer to $E_F$.
}
\end{figure}

Next, we consider the band results. The basic results from the simple, one-dimensional model, 
are confirmed by the band calculations, and the gap can be formed over the whole FS for varying doping. 
For the cases with AFM modulations, we first apply sinus-like modulations in the diagonal direction. The 
staggered fields (on Cu sites) have the same maximal amplitude as for the undoped insulating case. 
The undoped system contains 63 electrons per f.u.. However, the gaps (or pseudogaps) found in the four cases
with elongated cells are always found at one state (1 electron of each spin) below this electron
filling. This means at band fillings of 502, 628, 754 and 1006 electrons, for the four supercells of increasing lengths.
The doping is adjusted to coincide with these fillings, by use of the virtual crystal approximation. 
 The DOS near $E_F$ with and without AFM modulation, are shown in fig.
\ref{dos104} for the 104-site cell. The gap is small, like a pseudogap, and it coincides with $E_F$ 
when the doping is 0.25 holes per f.u., corresponding to the particular $q$. Modulations in the
longer cells give similar small gaps, or pseudogaps, but at positions corresponding to decreased hole doping,
at 0.2, 0.1667 and 0.125 holes per f.u., for the cells with 130, 156 and 208 sites, respectively. The opening of the
pseudogap starts along the diagonal $\Gamma-M$. Larger fields are needed to form a complete gap 
extending over the whole FS. Thus, in
agreement with the simple model, the
band results show that the barrel band can be cut to form gaps at different positions along $\Gamma - M$, depending
on the Fourier component of the potential perturbation.

AFM modulations in the x-direction is made for the 208-site cell, where the pseudogap is found for a
doping of 0.125 hole per f.u., as for the longest cell along the diagonal direction. 
By folding the k-points in the BZ back to the original BZ, it is seen that the partial opening of a gap
is largest far from the diagonal. This is consistent with the observations made by photoemission,
where the FS seems to "disappear" except near the diagonal \cite{norm}. (It is assumed that stripes are
running both along x and y \cite{oren}, making their effect superimposed.) Also the
observation of static stripes along [1,0,0] in one high-T$_c$ material \cite{tran} suggests 
that [1,0,0]-modulations are more probable than diagonal ones. 
This can be understood from the band results as follows: The larger z-dispersion
near off-diagonal points on the FS can be translated into higher DOS on the FS for these points, so that 
a band opening at those points will lead to a larger gain in total energy than if the gap opened along
the diagonal.
As the orientation in real space can be associated with the different FS points, it is therefore preferable
for the system to chose the [1,0,0]-orientation leading to the largest gain in energy.

These spin-polarized band calculations are not ab-initio, and they cannot determine the
precise form and amplitude of the modulations
in the doped cases. One can expect that some kind of
modulation is present, which in this band description produces gaps and stripes. Other theories
 based on strong on-site correlation have been shown to lead to stripe formation and spin-charge
 separation \cite{oren}.
Here, in a band description, the spin, charge and local DOS information show that 
the charges on Cu of different polarization are very similar.
The DOS at $E_F$ is largest on the non-polarized Cu-sites, but still sizable
on the polarized Cu sites, at least for one of the spins. Thus, conduction is not only
restricted to the zero-moment part of such stripes.

In the present model, there is a close connection between the doping and the wavelength of the stripe. The lower
the hole doping, the longer wavelengths. Further, it is probable that long wavelength modulations are more
stable than short ones. This is because at least two Cu sites have no moment in each modulation,
i.e. no exchange energy can be gained from a polarization of these two sites. The fraction of zero moment Cu sites 
is larger for short wavelength modulations, making them less stable. This cannot be verified by ab-initio
total energy calculations, but for the short modulation of only six Cu-layers along [1,0,0] it is 
more difficult to open a gap even for large field amplitudes. Hybridization between near neighbors 
in combination with rapid variations
of the potential over only a few layers will prevent an easy opening of a gap. In addition there is a topologic
change of the FS at the X-point when the doping approaches 0.25.
Therefore, a possibility is that pseudogaps, caused by spin waves, cannot exist for too large doping. This is
consistent with observations of pseudogaps only in underdoped high-T$_C$ oxides. A critical
concentration of about 0.19 has been suggested \cite{tallon}. This is to be compared with the values 0.125 for the 
shortest modulation along [1,0,0] and 0.25 along [1,1,0] to produce a "good" pseudogap in the calculations.
 
 The non-spinpolarized band calculations for phonon modes are ab-initio within LDA, 
 for the imposed lattice distortions.  Coupling between spin waves and phonons is plausible, but
we do not study such cases here. Different types of lattice distortions are considered, modulated along [1,0,0] and
[1,1,0] in the elongated supercells. Several distortions have a direct influence on the Cu potentials like
the change the distance between Cu and apical-O, or
between Ba and the CuO plane. The strongest effects occur for
"breathing" modes, where the four planar O sites are moved alternatively towards or away from the central Cu.
The modulations are chosen to be sinus-like.
Distortion amplitudes of the order 0.01$\cdot a$ are expected at low temperature for typical force constants 
and zero-point motions
\cite{san}, but in order to enhance the effects to be seen in the DOS, we also consider
amplitudes of 0.03$\cdot a$.

The DOS for the 104 site cell is shown in fig. \ref{dos104}. 
 For smaller distortion amplitudes the effect is smaller and hardly seen on the DOS. 
 By comparing the back-folded k-points with the points
in the normal BZ, one can verify that the effects due to modulations
along the diagonal or along the CuO-bonding, indeed are concentrated to the expected parts of the FS.
 Moreover, if the sinus-like modulation is replaced by the absolute value of sinus, one finds about
one order of magnitude smaller energy shifts near $E_F$.
 Results for the other cells are similar, so that the distorted structures
show some different DOS compared to the undistorted cases, but no real gaps are developed. However,
strong changes of the energies means that $\lambda$ is large for these modes. 
Full calculations
of $\lambda$ will not be presented here, but a few observations can be made. As was argued
above, the wavelengths are determined via doping so that only energies close
to $E_F$ are changing. 
Calculations of electron-phonon coupling made for the undoped compositions \cite{kra,san} will miss this effect, since the
sensitivity to $V_q$ of a certain wavelength appears below the position of $E_F$ for the undoped system. (The
wavelength corresponding to no doping would be infinite.)
 The largest change in energy
from the model above, is equal to $V_q$, i.e. equal to the maximal amplitude of a sinusoidal potential
perturbation. Two examples of O- and Ba-distortions with $V_q$ from the potential in the band calculation,
give $\lambda$ in the range 1-2. This is from
an approximate expression \cite{san} for the long-range 
coupling, $\lambda = N V_q^2 / \sum K (\Delta R)^2$,
where $N$ is the DOS at $E_F$ per cell. $K$ is a force constant \cite{san}, 
$\Delta R$ the displacement of each site, so that $K$ times
the sum over all sites gives the total elastic energy of the deformation. 
Electronic screening in a good metal tend to diminish the long-range part of $\lambda$, since  charge relaxation
close to a local lattice distortion will prevent any far-reaching change in Coulomb potential  \cite{sro}. 
As these materials are close to a metal-insulator transition, it is possible that the mechanism of screening
is reduced. The mechanism should retain the property of forward scattering of $\lambda$ which is favorable
to d-wave pairing \cite{weg}. Strong coupling for large wavelengths will renormalize
phonons near the zone center.
It is not yet clear if low or high frequency phonons will contribute most to superconductivity. 

In conclusion, the simple model of gap formation, supported by band results for HgBa$_2$CuO$_4$, suggests a
connection between spin waves, stripes, pseudogaps and strong electron-phonon coupling. The
reason is that the barrel band is sensitive 
to short-q perturbations, $V_q$. This scenario explains a number of
features of the high-$T_C$ oxides mentioned above.
 The origin of the pseudogaps is the stripe like spin-waves, which are stable only below a certain doping,
which in our model means above a certain wavelength of the spin modulation.
Whether they are dynamic 
spin-fluctuations, static or coupled to phonons, cannot yet be determined from the present calculations,
which are static.
 Energy considerations suggest 
that the spin waves are more probable at long wavelengths, i.e. in underdoped cases. 
Decreasing stability of the spin waves at short wavelength, i.e. for increasing hole doping, suggests that 
they and their role diminish with increasing doping.  
If this picture is true, and spinfluctuations
are harmful to superconductivity, one can expect that by applying pressure to samples with little doping 
it will be possible to suppress the fluctuations and reach higher $T_C$.
The effect of pressure should be smaller in overdoped samples.
 The band is equally
sensitive to perturbations $V_q$, produced by phonons. The amplitudes of $V_q$ due to phonons of realistic
deformation amplitudes at low T,
are smaller than in the case of AFM modulation, and fully developed gaps are normally not visible in the DOS. But
a particular $V_q$ induced by phonons favors  band shifts at some sections of the FS and a large $\lambda$.
Any perturbation, phonons, spin or valence fluctuations leading to a similar $V_q$ component will show
similar effects on the electronic states. Competition between different mechanisms 
can make interpretation of experiments difficult.






\begin{figure}[tb!] 
\leavevmode\begin{center}\epsfxsize8.6cm\epsfbox{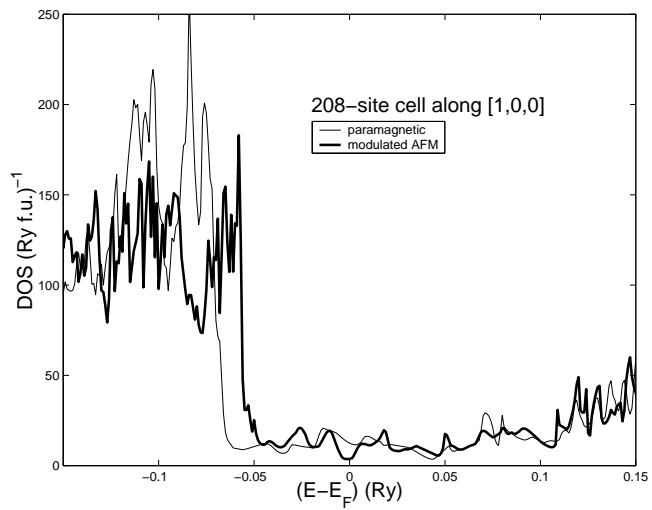}\end{center}
\caption{\label{dos208}
Paramagnetic (thin line) and AFM (heavy line) DOS for a 208-site unitcell of HBCO 
(doped with 0.125 holes per f.u.) along [1,0,0].
}
\end{figure}


\end{document}